\documentclass[12pt]{article}%
\usepackage{graphicx}
\usepackage{amsmath}
\usepackage{epsfig}
\usepackage{amsfonts}
\usepackage{amssymb}
\usepackage[cp866]{inputenc}
\usepackage[T2A]{fontenc}
\usepackage[english]{babel}%
\providecommand{\U}[1]{\protect\rule{.1in}{.1in}}
\providecommand{\U}[1]{\protect\rule{.1in}{.1in}}
\begin{document}

\title{Polarization operator of a photon in a magnetic field}
\author{V.M. Katkov\\Budker Institute of Nuclear Physics,\\Novosibirsk, 630090, Russia\\e-mail: katkov@inp.nsk.su}
\maketitle

\begin{abstract}
In the first order of $\alpha,$ the polarization operator of a photon is
investigated in a constant and homogeneous magnetic field at arbitrary photon
energies. For weak and strong fields $H$ (compared with the critical field
$H_{0}=\ 4.41\cdot10^{13}%
\operatorname{G}%
),$ approximate expressions have been found. We consider the pure quantum
region of photon energy near the threshold of pair creation, as well as the
region of high energy levels where the quasiclassical approximation is valid.
The general formula has been obtained for the effective mass of photon with
given polarization. It is useful for an analysis of the problem under
consideration on the whole and at a numerical work

\end{abstract}

\section{Introduction}

The study of QED processes in a strong magnetic field close to and exceeding
the critical field strength $H_{0}=m^{2}/e=4,41\cdot10^{13}$ $%
\operatorname{G}%
$ (the system of units $\hbar=c=1$ is used ) is stimulated essentially by the
existence of very strong magnetic field in nature. It is universally
recognized the magnetic field of neutron stars (pulsars) run up $Н \sim
10^{11}\div10^{13}%
\operatorname{G}%
$ \cite{[1]}. These values of field strength gives the rotating magnetic
dipole model, in which the pulsar loses rotational energy through the magnetic
dipole radiation. The prediction of this model is in quite good agreement with
the observed radiation from pulsars in the radio frequency region. There are
around some thousand radio pulsars. Another class of neutron stars, now
referred to as magnetars \cite{[2]}, was discovered on examination of the
observed radiation at x-ray and $\gamma$-ray energies and may possess even
stronger surface magnetic fields $Н \sim10^{14}\div10^{15}$ $%
\operatorname{G}%
$. The photon propagation in these fields and the dispersive properties of the
space region with magnetic is of very much interest. This propagation
accompanied by the photon conversion into a pair of charged particles when the
transverse photon momentum is larger than the process threshold value
$k_{\perp}>2m.$ When the field change is small on the characteristic length of
process formation (for example, when this length is smaller then the scale of
heterogeneity of the neutron star magnetic field), the consideration can be
realized in the constant field approximation. In 1971 Adler \cite{[3]} had
calculated the photon polarization operator in a magnetic field using the
proper-time technique developed by Schwinger \cite{[4]}. In the same year
Batalin and Shabad \cite{[5]} had calculated this operator in an
electromagnetic field using the Green function found by Schwinger \cite{[4]}.
In 1975 the contribution of charged-particles loop in an electromagnetic field
with $n$ external photon lines had been calculated in \cite{[6]}. For $n=2$
the explicit expressions for the contribution of scalar and spinor particles
to the polarization operator of photon were given in this work. For the
contribution of spinor particles obtained expressions coincide with the result
of \cite{[5]}, but another form is used.

The polarization operator in a constant magnetic field was investigated well
enough in the energy region lower and near the pair creation threshold (see,
for example, the papers \cite{[7], [8], [9]} and the bibliography cited there.
In the present paper we consider in detail the polarization operator on mass
shell ( $k^{2}=0,$the metric $ab=$ $a^{0}b^{0}-\mathbf{ab}$ is used ) at
arbitrary value of the photon energy and magnetic field strength. The
restriction of our consideration is only the applicability of the perturbation
theory over the electromagnetic interaction constant $\alpha.$

\section{General expressions for the polarization operator}

Our analysis is based on the general expression for the contribution of spinor
particles to the polarization operator obtained in a diagonal form in
\cite{[6]} (see Eqs. (3.19), (3.33)). For the case of pure magnetic field we
have in a covariant form the following expression%

\begin{align}
\Pi^{\mu\nu}  &  =-\sum_{i=2,3}\kappa_{i}\beta_{i}^{\mu}\beta_{i}^{\nu
},\ \ \ \beta_{i}\beta_{j}=-\ \delta_{ij},\ \ \ \beta_{i}k=0;\label{1}\\
\beta_{2}^{\mu}  &  =(F^{\ast}k)^{\mu}/\sqrt{-(F^{\ast}k)^{2}},\ \ \ \beta
_{3}^{\mu}=(Fk)^{\mu}/\sqrt{-(F^{\ast}k)^{2}},\ \nonumber\\
\ FF^{\ast}  &  =0,\ \ \ F^{2}=F^{\mu\nu}F_{\mu\nu}=2(H^{2}-E^{2})>0,
\label{01}%
\end{align}
where $F^{\mu\nu}-$ the electromagnetic field tensor , $F^{\ast\mu\nu}-$ dual
tensor, $k^{\mu}$ $-$ the photon momentum, $(Fk)^{\mu}=F^{\mu\nu}k_{\nu},$%

\begin{equation}
\kappa_{i}=\frac{\alpha}{\pi}m^{2}r%
{\textstyle\int\limits_{-1}^{1}}
dv%
{\textstyle\int\limits_{0}^{\infty-\mathrm{i0}}}
f_{i}(v,x)\exp[\mathrm{i}\psi(v,x)]dx. \label{2}%
\end{equation}
Here%

\begin{align}
f_{2}(v,x)  &  =2\frac{\cos(vx)-\cos x}{\sin^{3}x}-\frac{\cos(vx)}{\sin
x}+v\frac{\cos x\sin(vx)}{\sin^{2}x},\nonumber\\
f_{3}(v,x)  &  =\frac{\cos(vx)}{\sin x}-v\frac{\cos x\sin(vx)}{\sin^{2}%
x}-(1-v^{2})\cot x,\nonumber\\
\psi(v,x)  &  =\frac{1}{\mu}\left\{  2r\frac{\cos x-\cos(vx)}{\sin
x}+[r(1-v^{2})-1]x\right\}  ;\label{3}\\
r  &  =-(F^{\ast}k)^{2}/2m^{2}F^{2},\ \ \ \mu^{2}=F^{2}/2H_{0}^{2}. \label{03}%
\end{align}
The real part of $\kappa_{i}$ determines the refractive index $n_{i}$ of the
photon with polarization $e_{i}=\beta_{i}$:%

\begin{equation}
\ \ \ \ n_{i}=1-\frac{\mathrm{\operatorname{Re}}\kappa_{i}}{2\omega^{2}}.
\label{4}%
\end{equation}
At $r>1$ the proper value of polarization operator $\kappa_{i}$ includes the
imaginary part which determines the probability per unit length of pair
production by photon with the polarization $\beta_{i}$:%

\begin{equation}
W_{i}=-\frac{1}{\omega}\mathrm{\operatorname{Im}}\kappa_{i} \label{5}%
\end{equation}
For $r<1$ the integration counter over $x$ in Eq. (\ref{2}) may be turn to the
lower axis $(x\rightarrow-\mathrm{i}x),$ then the value $\kappa_{i}$ becomes
real in the explicit form.

\section{Weak field and low energy: $\mu\ll1,$ $1<r\ll1/\mu^{2}$}

Let's remove the integration counter over $x$ in Eq. (\ref{2}) to the lower
axis at the value $x_{0}:$%

\begin{equation}
x_{0}(r)=-\mathrm{i}l(r),\ \ \ \ l(r)=\ln\frac{\sqrt{r}+1}{\sqrt{r}-1}.
\label{6}%
\end{equation}
As a result we have the following expression for $\kappa_{i}:$%

\begin{equation}
\kappa_{i}=\frac{\alpha}{\pi}m^{2}r\left(  a_{i}+b_{i}\right)  , \label{06}%
\end{equation}
where%

\begin{align}
a_{i}  &  =-\mathrm{i}%
{\textstyle\int\limits_{-1}^{1}}
dv%
{\textstyle\int\limits_{0}^{l(r)}}
dxf_{i}(v,-\mathrm{i}x)\exp[\mathrm{i}\psi(v,-\mathrm{i}x)],\label{7}\\
b_{i}  &  =%
{\textstyle\int\limits_{-1}^{1}}
dv%
{\textstyle\int\limits_{0}^{\infty}}
dzf_{i}(v,z+x_{0})\exp[\mathrm{i}\psi(v,z+x_{0})]. \label{07}%
\end{align}
In the integral $a_{i}$ in Eq. (\ref{7}) the small values $x\sim\mu$
contribute. This integral we calculate expanding the entering functions over
$x.$ Taking into account that in the region under consideration the condition
$r\mu^{2}\ll1$ is fulfilled we keep in the exponent argument the term $-x/\mu$
only and extend the integration over $x$ to infinity. In the result of not
complicated integration over $v$ we have:%

\begin{align}
a_{2}  &  =-\frac{16}{45}\mu^{2},\ \ \ a_{3}=-\frac{28}{45}\mu^{2}%
;\ \nonumber\\
\ \ \kappa_{2}^{a}  &  =-\frac{4\alpha m^{2}\kappa^{2}}{45\pi},\ \ \ \kappa
_{3}^{a}=-\frac{7\alpha m^{2}\kappa^{2}}{45\pi},\ \ \ \ \kappa^{2}%
=-\frac{(Fk)^{2}}{H_{0}^{2}m^{2}}. \label{8}%
\end{align}
In the integral $b_{i}$ Eq. (\ref{07}) the small values $v$ contribute.
Expanding entering functions over $v$ and extending the integration over $v$
to infinity we have%

\begin{align}
b_{i}  &  =\sqrt{\mu\pi}\exp\left(  -\mathrm{i}\frac{\pi}{4}\right)
{\displaystyle\int\limits_{0}^{\infty}}
\frac{dzf_{i}(0,x_{0}+z)}{\sqrt{\chi(x_{0}+z)}}\exp\left[  -\frac{\mathrm{i}%
}{\mu}\varphi(x_{0}+z)\right]  ,\label{9}\\
\varphi(x)  &  =2r\tan\left(  \frac{x}{2}\right)  +(1-r)x,\ \ \ \chi
(x)=rx\left(  1-\frac{x}{\sin x}\right)  .\nonumber
\end{align}

We consider now the energy region where $r-1\ll1$ when the moving of created
particles is nonrelativistic. In this case%

\begin{align}
\mathrm{i}\varphi(z+x_{0})/\mu &  \simeq\beta(r)-\gamma(e^{-\mathrm{i}%
z}+\mathrm{i}z),\ \chi(x)\simeq z+x_{0};\nonumber\\
\ \ \ \beta(r)  &  =2\sqrt{r}/\mu+\gamma(1-l(r)),\ \gamma=(r-1)/\mu
,\ \nonumber\\
\ \ \ f_{2}(0,x)  &  \simeq0,\ \ \ f_{3}(0,x)\simeq-\mathrm{i.}. \label{10}%
\end{align}
In the threshold region ($\gamma\sim1),$ where the particles occupy not very
high energy levels, we present Eq. (\ref{9}) for $b_{3}$ in the form%

\begin{align}
b_{3}  &  =-\mathrm{i}\sqrt{\mu\pi}\exp\left(  -\mathrm{i}\frac{\pi}{4}%
-\beta(r)\right) \nonumber\\
&  \times%
{\textstyle\int\limits_{0}^{\infty}}
\frac{dz}{\sqrt{x_{0}+z}}%
{\textstyle\sum\limits_{n=0}^{\infty}}
\frac{\gamma^{n}}{n!}\exp[\mathrm{i}(\gamma-n)z],\ \label{11}%
\end{align}
The integral in Eq. (\ref{11}) has a root singularity at whole numbers of
$\gamma=k.$ For $|\gamma-k|\ \ll x_{0}^{-1}$, $z\sim$ $|\gamma-k|^{-1}>>x_{0}$
we have:%

\begin{align}
b_{3}  &  =-2\mathrm{i}\sqrt{\mu\pi}\exp[-\mathrm{i}\frac{\pi}{4}%
-\beta(r)]\frac{\gamma^{k}}{k!}%
{\textstyle\int\limits_{0}^{\infty}}
dy\exp[\mathrm{i}(\gamma-k)y^{2}]\nonumber\\
=  &  -\pi\sqrt{\frac{\mu}{|\gamma-k|}}\exp\left[  -\beta(r)+\mathrm{i}%
\frac{\pi}{2}\vartheta(\gamma-k)\right]  \frac{k^{k}}{k!}, \label{12}%
\end{align}
where $\vartheta(z)-$ Heaviside function: $\vartheta(z)=1$ for $z\geqslant0,$
$\vartheta(z)=0$ for $z<0.$ The expression for $\kappa_{3}$ with the accepted
accuracy can be rewritten in the following form:%

\begin{align}
\kappa_{3}^{b}  &  \simeq-\alpha m^{2}\mu e^{-\zeta}%
{\textstyle\sum\limits_{n=0}^{\infty}}
\frac{(2\zeta)^{n}}{\sqrt{|g|}n!}\exp\left[  \mathrm{i}\frac{\pi}{2}%
\vartheta(g)\right]  ,\ \nonumber\\
\ g  &  =r-1-n\mu,\ \ \zeta=2r/\mu. \label{13}%
\end{align}

At $\gamma\gg1$ the small $z$ contributes to the integral in Eq. (\ref{9})$,$ then:%

\begin{align}
\mathrm{i}\varphi(z+x_{0})/\mu &  \simeq p(r)+\gamma z^{2}/2,\ \ \mathrm{i}%
\chi(z+x_{0})\ \simeq\mathrm{i}x_{0}=l(r),\nonumber\\
\ \ p(r)  &  =\frac{2\sqrt{r}}{\mu}-\gamma l(r),\ \ \gamma=\frac{r-1}{\mu
}\ ,\ \ l(r)=\ln\frac{\sqrt{r}+1}{\sqrt{r}-1}. \label{14}%
\end{align}
In the result of simple integration over $z$ we have%

\begin{equation}
\kappa_{3}^{b}\simeq-\mathrm{i}\alpha m^{2}e^{-p(r)}\sqrt{\frac{\mu}{2\gamma
l(r)}},\ \ \kappa_{3}^{b}\simeq0. \label{140}%
\end{equation}
In a very wide range of energies, when the condition $1\lesssim r-1\ll\mu
^{-2}$ is fulfilled, in Eq. (\ref{07}) for $b_{i}$ one can carry out the
expansion over $v$ and $z$ from the very outset. As a result we have (see
\cite{[10]}, Eq. (B5)):%
\begin{equation}
\kappa_{3}^{b}\simeq-\mathrm{i}\alpha m^{2}e^{-p(r)}\sqrt{\frac{r}{\gamma
l(r)p(r)}},\ \ \ \kappa_{2}^{b}=\frac{r-1}{2r}\kappa_{3}^{b}. \label{014}%
\end{equation}

At $r-1\ll1,$ the last equation for $\kappa_{i}$ is consistent with the
previous expression.

\section{ Weak field and high energy: $\mu\ll1,$ $r\gtrsim1/\mu^{2}$}

This region is contained in the region of the standard quasiclassical
approximation (SQA) \cite{[10]}, \cite{[11]}.\ The main contribution to the
integral in Eq. (\ref{2}) is given by small values of $x.$ Expanding the
entering functions Eq. (\ref{3}) over $x$, and carrying out the change of
variable $x=\mu t,$ we get%

\begin{align}
\kappa_{i}  &  =\frac{\alpha m^{2}\kappa^{2}}{24\pi}%
{\textstyle\int\limits_{0}^{1}}
\alpha_{i}(v)(1-v^{2})dv%
{\textstyle\int\limits_{0}^{\infty}}
t\exp[-\mathrm{i}(t+\xi\frac{t^{3}}{3})]dt;\label{15}\\
\alpha_{2}  &  =3+v^{2},\ \ \alpha_{3}=2(3-v^{2}),\ \nonumber\\
\ \sqrt{\xi}  &  =\frac{\kappa(1-v^{2})}{4},\ \ \kappa^{2}=4r\mu^{2}%
=-\frac{(Fk)^{2}}{m^{2}H_{0}^{2}}. \label{150}%
\end{align}
The entering in Eq. (\ref{15}) integrals over $t$ are expressed by derivations
of Airy (the imaginary part) and Hardy (the real part) integrals. Because of
the application conditions, this energy region is overlapped with the
considered above. At $\kappa\ll1$ we have for the integrals entering into the
real part of $\kappa_{i}$%

\begin{equation}%
{\textstyle\int\limits_{0}^{\infty}}
t\cos tdt=-1,\ \ \
{\textstyle\int\limits_{0}^{1}}
\alpha_{2}(1-v^{2})dv=\frac{32}{15},\ \ \ \
{\textstyle\int\limits_{0}^{1}}
\alpha_{3}(1-v^{2})dv=\frac{56}{15}. \label{16}%
\end{equation}
These expression coincides with (\ref{8}).

At calculation of the imaginary part of the integral over $t$ in Eq.
(\ref{15}) we extend the integration to the whole axis because of the
integrand parity. After that, the stationary phase method can be used (
$t_{0}=-$ $\mathrm{i}$ $/\sqrt{\xi}$ ). As a result of the standard procedure
of above method we have%

\begin{align}
&  \frac{1}{2}%
{\textstyle\int\limits_{-\infty}^{\infty}}
t\exp[-\mathrm{i}(t+\xi\frac{t^{3}}{3})]dt\nonumber\\
&  =\frac{t_{0}}{2}\sqrt{\frac{\pi}{\mathrm{i}t_{0}\xi}}\exp[-\mathrm{i}%
(t_{0}+\xi\frac{t_{0}^{3}}{3})]\nonumber\\
&  =-\mathrm{i}\frac{\sqrt{\pi}}{2}\xi^{-3/4}\exp\left(  -\frac{2}{3\sqrt{\xi
}}\right)  =\nonumber\\
&  -4\mathrm{i}\sqrt{\pi}\kappa^{-3/2}\left(  1-v^{2}\right)  ^{-3/2}%
\exp\left(  -\frac{8}{3\kappa(1-v^{2})}\right)  . \label{170}%
\end{align}
Substituting the obtained expression into Eq. (\ref{15}) and fulfilling the
integration over $v,$ keeping in mind the small $v$ contributes, we get:%

\begin{equation}
\kappa_{2}=-\mathrm{i}\sqrt{\frac{3}{32}}\alpha m^{2}\kappa\exp\left(
-\frac{8}{3\kappa}\right)  ,\ \ \ \kappa_{3}=2\kappa_{2}. \label{18}%
\end{equation}
At $\kappa\gg1$ ($\xi\gg1$) the small $t$ contributes to the integral
(\ref{15}) ( $\xi t^{3}\sim1),$ and in the argument of exponent Eq. (\ref{15})
the linear over $t$ term may be omit. Carrying out the change of variable :%

\begin{equation}
\xi t^{3}/3=-\mathrm{i}x,\ \ \ t=\exp\left(  \frac{-\mathrm{i}\pi}{6}\right)
\left(  \frac{3x}{\xi}\right)  ^{1/3}, \label{19}%
\end{equation}
one obtains:%

\begin{equation}
\kappa_{i}=\frac{\alpha m^{2}\kappa^{2}}{24\pi}\exp\left(  \frac
{-\mathrm{i}\pi}{3}\right)  \frac{1}{3}\left(  \frac{48}{\kappa^{2}}\right)
^{2/3}\Gamma\left(  \frac{2}{3}\right)
{\textstyle\int\limits_{0}^{1}}
dv\alpha_{i}(v)(1-v^{2})^{-1/3}. \label{20}%
\end{equation}
After integration over $v$ we have:%

\begin{equation}
\kappa_{i}=\frac{\alpha m^{2}(3\kappa)^{2/3}}{7\pi}\frac{\Gamma^{3}\left(
\frac{2}{3}\right)  }{\Gamma\left(  \frac{1}{3}\right)  }(1-\mathrm{i}\sqrt
{3})\beta_{i},\ \ \ \beta_{2}=2,\ \ \ \beta_{3}=3. \label{21}%
\end{equation}

\section{ Strong fields: $\mu\gtrsim1$}

Let's consider the energy region the upper boundary of which is slightly
higher the characteristic energy $r_{10}$. At this energy one of the particles
is created on the first excited level and another particle is created on the
lower level. At that we choose the lower boundary of the energy region
slightly below the threshold energy $r_{00}$:%

\begin{align}
\ \ r_{lk}  &  =(\varepsilon(l)+\varepsilon(k))^{2}/4m^{2},\nonumber\\
\ \ \ \varepsilon(l)  &  =\sqrt{m^{2}+2eHl}=m\sqrt{1+2\mu l}.\ \ \ \label{021}%
\end{align}
For $r<r_{10},$ the integration counter over $x$ in Eq. (\ref{2}) may be turn
to the lower imaginary axis for all terms in the integrand except the term in
expression for $\kappa_{3},$ which contains the function $\Phi(v,x)$ $=$
$-(1-v^{2})\mathrm{ctg}x\exp[\mathrm{i}\psi(v,x)]$. Let's add to $\Phi(v,x)$
and take off the function%
\begin{align}
\Phi_{_{\mathrm{red}}}(v,x)  &  =\mathrm{i}(1-v^{2})\exp[\mathrm{i}%
\psi_{\mathrm{red}}(v,x)],\ \ \ \nonumber\\
\ \ \psi_{\mathrm{red}}(v,x)  &  =\frac{1}{\mu}\left\{  2r\mathrm{i}%
+[r(1-v^{2})-1]x\right\}  . \label{22}%
\end{align}
After that in the integral over $x$ for the sum $\Phi(v,x)+\Phi
_{_{\mathrm{red}}}(v,x),$ the integration counter over $x$ can be turn to the
lower axis. The integral over $x$ for the residuary function has the the
following form%

\begin{align}
&
{\textstyle\int\limits_{0}^{\infty}}
\exp[\mathrm{i}\psi_{\mathrm{red}}(v,x)]dx\nonumber\\
&  =\exp\left(  -\frac{2r}{\mu}\right)  \frac{\mathrm{i}\mu}{r(1-v^{2}%
)-1+\mathrm{i}0}\nonumber\\
&  =\mu\exp\left(  -\frac{2r}{\mu}\right)  \left[  \mathrm{i}\frac
{\text{\textit{ }}\mathcal{P}}{r-1-rv^{2}}+\pi\delta\left(  r-1-rv^{2}\right)
\right]  . \label{23}%
\end{align}
The operator $\mathcal{P}$ means the principal value integral. Carrying out
the integration over $v,$ we have%

\begin{align}
-\mathrm{i}r  &
{\textstyle\int\limits_{-1}^{1}}
dv(1-v^{2})\left[  \mathrm{i}\frac{\text{\textit{ }}\mathcal{P}}{r-1-rv^{2}%
}+\pi\delta\left(  r-1-rv^{2}\right)  \right] \nonumber\\
&  =2\left[  1+\frac{1}{\sqrt{r(1-r)}}\mathrm{\arctan}\sqrt{\frac{1-r}{r}%
}\right]  -\frac{\pi}{\sqrt{r(1-r)}}. \label{24}%
\end{align}
Finally the expression for $\kappa_{i}$ takes the following well-behaved form:%

\begin{align}
\kappa_{2}  &  =\alpha m^{2}\frac{r}{\pi}%
{\textstyle\int\limits_{-1}^{1}}
dv%
{\textstyle\int\limits_{0}^{\infty}}
F_{2}(v,x)\exp[-\chi(v,x)]dx,\ \ \ \kappa_{3}=\kappa_{3}^{1}+\kappa_{3}%
^{00},\label{124}\\
\kappa_{3}^{1}  &  =\alpha m^{2}\frac{r}{\pi}%
{\textstyle\int\limits_{-1}^{1}}
dv%
{\textstyle\int\limits_{0}^{\infty}}
\{F_{3}(v,x)\exp[-\chi(v,x)]\nonumber\\
&  +(1-v^{2})\exp[-\chi_{00}(v,x)]dx\},\label{025}\\
\kappa_{3}^{00}  &  =\alpha m^{2}\frac{\mu}{\pi}\exp\left(  -\frac{2r}{\mu
}\right)  \left[  2+B(r)\right]  ;\ \label{25}\\
\ B(r)  &  =\frac{2}{\sqrt{r(1-r)}}\arctan\sqrt{\frac{1-r}{r}}-\frac{\pi
}{\sqrt{r(1-r)}}\ . \label{125}%
\end{align}
At the photon energy higher threshold ( $r>1,$ $\sqrt{1-r}=-\mathrm{i}%
\sqrt{r-1}$):%

\begin{equation}
B(r)=\frac{2}{\sqrt{r(r-1)}}\ln(\sqrt{r}+\sqrt{r-1})-\frac{\mathrm{i}\pi
}{\sqrt{r(r-1)}}. \label{026}%
\end{equation}
Here%

\begin{align}
F_{2}(v,x)  &  =\frac{1}{\sinh x}\left(  2\frac{\cosh x-\cosh(vx)}{\sinh^{2}%
x}-\cosh(vx)+v\sinh(vx)\coth x\right)  ,\nonumber\\
F_{3}(v,x)  &  =\frac{\cosh(vx)}{\sinh x}-v\frac{\cosh x\sinh(vx)}{\sinh^{2}%
x}-(1-v^{2})\coth x;\label{26}\\
\chi(v,x)  &  =\frac{1}{\mu}\left[  2r\frac{\cosh x-\cosh(vx)}{\sinh
x}+(rv^{2}-r+1)x\right]  ,\label{027}\\
\chi_{00}(v,x)  &  =\frac{1}{\mu}\left[  2r+(rv^{2}-r+1)x\right]  . \label{27}%
\end{align}
For superstrong fields ($\mu\gg1),$ the entering into integrands of Eqs.
(\ref{124}) and (\ref{025}) exponential terms can be substitute for unit. As a
result we have for leading terms%

\begin{equation}
\kappa_{2}\simeq-\frac{4r}{3\pi}\alpha m^{2},\ \ \ \ \kappa_{3}\simeq-\alpha
m^{2}\frac{\mu}{\pi}(2+B(r)). \label{028}%
\end{equation}

The integrals for $\kappa_{2}$ and $\kappa_{3}^{1}$ have the root divergence
at $r=r_{10}.$To bring out these distinctions in an explicit form, let's
consider the main asymptotic terms of corresponding integrand at
$x\rightarrow\infty$:%

\begin{align}
\kappa_{i}^{10}  &  =\alpha m^{2}r\frac{2}{\pi}%
{\textstyle\int\limits_{-1}^{1}}
dv%
{\textstyle\int\limits_{0}^{\infty}}
d_{i}(v)\exp[-\chi_{10}(v,x)]dx,\ \label{28}\\
\ \ d_{2}  &  =v-1,\ d_{3}=1-v-\frac{2r}{\mu}(1-v^{2})\label{280}\\
\chi_{10}(v,x)  &  =\frac{2r}{\mu}+\frac{1}{\mu}\left[  (1-v)\mu
+rv^{2}-r+1\right]  x. \label{29}%
\end{align}
After elementary integration over $x,$ one gets%
\begin{equation}
\kappa_{i}^{10}=\alpha m^{2}\mu r\frac{2}{\pi}\exp\left(  -\frac{2r}{\mu
}\right)
{\textstyle\int\limits_{-1}^{1}}
dv\frac{d_{i}(v)}{rv^{2}-\mu v-r+1+\mu}. \label{30}%
\end{equation}
Performing integration over $v$, we have:%

\begin{align}
\kappa_{2}^{10}  &  =\alpha m^{2}\mu r\frac{2}{\pi}\exp\left(  -\frac{2r}{\mu
}\right)  \left[  \frac{\mu/2r-1}{\sqrt{h(r)}}A(r)-\frac{1}{2r}\ln
(2\mu+1)\right]  ,\label{31}\\
\kappa_{3}^{10}  &  =\alpha m^{2}\mu r\frac{2}{\pi}\exp\left(  -\frac{2r}{\mu
}\right) \nonumber\\
&  \times\left[  \frac{\mu/2r-1-2/\mu}{\sqrt{h(r)}}A(r)-\frac{1}{2r}\ln
(2\mu+1)+\frac{2}{\mu}\right]  ,\label{32}\\
A(r)  &  =\arctan\frac{r-\mu/2}{\sqrt{h(r)}}+\arctan\frac{r+\mu/2}{\sqrt
{h(r)}}\nonumber\\
&  =\pi-\arctan\frac{\sqrt{h(r)}}{r-\mu/2}-\arctan\frac{\sqrt{h(r)}}{r+\mu
/2},\label{33}\\
h(r)  &  =(1+\mu)r-r^{2}-\mu^{2}/4. \label{034}%
\end{align}
At $r=r_{10}=(1+\mu+\sqrt{1+2\mu})/2,$ $h(r)=0$ and expressions in Eqs.
(\ref{31})-(\ref{32}) contain the root divergence :%

\begin{equation}
\kappa_{i}^{10}\simeq-4\alpha m^{2}r\exp\left(  -\frac{2r}{\mu}\right)
\frac{\beta_{i}}{\sqrt{h(r)}},\ \ \beta_{2}=\frac{\mu}{2}-\frac{\mu^{2}}%
{4r},\ \ \beta_{3}=1+\frac{\mu}{2}-\frac{\mu^{2}}{4r}. \label{34}%
\end{equation}
For the higher photon energy $r>r_{10}$ (but $r<$ $r_{20}$ $=(1+\sqrt{1+4\mu
})^{2}/4$), the new channel of pair creation arises, and Eq. (\ref{30})
changes over (cf. (\ref{24})):%

\begin{align}
\kappa_{i}^{10}  &  =\alpha m^{2}\mu r\frac{2}{\pi}\exp\left(  -\frac{2r}{\mu
}\right) \nonumber\\
&  \times%
{\textstyle\int\limits_{-1}^{1}}
dvd_{i}(v)[\frac{\mathcal{P}}{rv^{2}-\mu v-r+1+\mu}\nonumber\\
&  -\mathrm{i}\pi\delta(rv^{2}-\mu v-r+1+\mu)]; \label{35}%
\end{align}
At $r-r_{10}<<1$%

\begin{equation}
\kappa_{i}^{10}\simeq-4\mathrm{i}\alpha m^{2}r\exp\left(  -\frac{2r}{\mu
}\right)  \frac{\beta_{i}}{\sqrt{-h(r)}}. \label{36}%
\end{equation}
This direct procedure of divergence elimination can be extended further. But
we consider, in the next section, another technique allowing to perform done
extracting in general case.

For strong fields and high energy levels ($\mu\gtrsim1,\ r\gg\mu$), Eqs.
(\ref{19}--\ref{21}) can be used because of $x\sim(\mu/r)^{1/3}<<1$
contributes, and condition $\kappa>>1$ is identically valid in this case.
Formula (\ref{21}) coincides with the corresponding formula of SQA at
$\kappa\gg1.$ However, be aware that for weak fields ($\mu\ll1,\ H\ll H_{0}),$
condition $\kappa\gg1$ is sufficient for the quasiclassical motion $(n\gg1)$
of produced particles. While for the fields significantly larger than the
critical field ($\mu\gg1)$, a large value of the parameter $\kappa$ does not
provide specified quasiclassicality. In this case, a prerequisite for the
applicability of the SQA is the condition $r/\mu\sim n\gg1.$

\section{General case}

As well as in our work \cite{[10]} (see Appendix A), we present the effective
mass in the form of%

\begin{align}
\kappa_{i}  &  =\alpha m^{2}\frac{r}{\pi}T_{i};\ \ T_{i}=%
{\textstyle\int\limits_{-1}^{1}}
dv%
{\textstyle\int\limits_{0}^{\infty-\mathrm{i0}}}
f_{i}(v,x)\exp[\mathrm{i}\psi(v,x)]dx,\ \ \label{37}\\
T_{i}  &  =%
{\displaystyle\sum\limits_{n=0}^{\infty}}
\left(  1-\frac{\delta_{n0}}{2}\right)  T_{i}^{(n)};\ \label{38}\\
\ \ T_{i}^{(n)}  &  =%
{\textstyle\int\limits_{-1}^{1}}
dv%
{\textstyle\int\limits_{0}^{\infty-\mathrm{i0}}}
F_{i}^{(n)}(v,x)\exp[\mathrm{i}a_{n}(v)x]dx, \label{380}%
\end{align}
where%

\begin{align}
F_{1}^{(n)}  &  =(-\mathrm{i})^{n}\exp(\mathrm{i}z\cot x)\left[
\frac{\mathrm{i}}{\sin x}(J_{n+1}(t)-J_{n-1}(t))-\frac{2vn}{z}\cot
xJ_{n}(t)\right]  ,\nonumber\\
F_{2}^{(n)}  &  =(-\mathrm{i})^{n}\exp(\mathrm{i}z\cot x)\frac{4}{z}\left(
b\cot x-\frac{\mathrm{i}}{\sin^{2}x}\right)  J_{n}(t)-F_{1}^{(n)},\nonumber\\
F_{3}^{(n)}  &  =F_{1}^{(n)}-2(-\mathrm{i})^{n}\exp(\mathrm{i}z\cot
x)(1-v^{2})\cot xJ_{n}(t);\label{39}\\
a_{n}(v)  &  =nv-b,\ \ b=\frac{1}{\mu}(1-r(1-v^{2})),\ \ z=\frac{2r}{\mu
},\ \ t=\frac{z}{\sin x}. \label{40}%
\end{align}
Let's note that at $x\rightarrow-\mathrm{i}\infty$ the asymptotic of the
Bessel function $J_{n}(t)$ is%

\begin{equation}
J_{n}(t)\simeq J_{n}(2\mathrm{i}ze^{-|x|})\simeq\frac{(\mathrm{i}z)^{n}}%
{n!}e^{-n|x|}, \label{41}%
\end{equation}
and under the condition $a_{n}(v)<n,$ the integration counter over $x$ in Eq.
(\ref{38}) can be unrolled to the lower axis. Then $T_{i}^{(n)}$ becomes real
in the explicit form.

The functions $F_{i}^{(n)}(v,x)$ are periodical over $x.$ So one can present
$T_{i}^{(n)}$as%

\begin{align}
T_{i}^{(n)}  &  =%
{\textstyle\int\limits_{-1}^{1}}
dv%
{\textstyle\int\limits_{0}^{2\pi}}
F_{i}^{(n)}(v,x)\exp[\mathrm{i}a_{n}(v)x]dx%
{\displaystyle\sum\limits_{k=0}^{\infty}}
\exp[2\pi\mathrm{i}ka_{n}(v)]\nonumber\\
&  =%
{\textstyle\int\limits_{-1}^{1}}
\frac{dv}{1-\exp[2\pi\mathrm{i}a_{n}(v)]+\mathrm{i}0}%
{\textstyle\int\limits_{0}^{2\pi}}
F_{i}^{(n)}(v,x)\exp[\mathrm{i}a_{n}(v)x]dx. \label{42}%
\end{align}
We use the well-known expression%

\begin{align}
&  \frac{1}{1-\exp[2\pi\mathrm{i}a_{n}(v)]+\mathrm{i}0}\nonumber\\
&  =\frac{\mathcal{P}}{1-\exp[2\pi\mathrm{i}a_{n}(v)]}-\mathrm{i}\pi
\delta(1-\exp[2\pi\mathrm{i}a_{n}(v)]), \label{43}%
\end{align}
Taking into account the above notation (\ref{41}), we have%

\begin{align}
&  -\mathrm{i}\pi\delta(1-\exp[2\pi\mathrm{i}a_{n}(v)])\nonumber\\
&  =-\mathrm{i}\pi%
{\displaystyle\sum\limits_{m}}
\delta(1-\exp[2\pi\mathrm{i}(a_{n}(v)-m)])\label{44}\\
&  \rightarrow\frac{1}{2}%
{\displaystyle\sum\limits_{m\geq n}}
\delta(a_{n}(v)-m).
\end{align}
Also using the ratio$\ \ \ \ \ \ F_{i}^{(n)}(v,x+\pi)=(-1)^{n}F_{i}%
^{(n)}(v,x),$ \ \ \ \ \ \ \ we get%

\begin{align}
T_{i}^{(n)}  &  =(-1)^{n}\frac{\mathrm{i}}{2}\mathcal{P}%
{\textstyle\int\limits_{-1}^{1}}
\frac{dv}{\sin(\pi a_{n}(v))}%
{\displaystyle\int\limits_{-\pi}^{\pi}}
F_{i}^{(n)}(v,x)\exp[\mathrm{i}a_{n}(v)x]dx\nonumber\\
&  +%
{\displaystyle\sum\limits_{m\geq n}^{m=n_{\max}}}
{\displaystyle\sum\limits_{v_{1,2}}}
\frac{1+(-1)^{m+n}}{2|a_{n}^{\prime}(v)|}\vartheta(g(n,m,r))\label{45}\\
&  \times%
{\displaystyle\int\limits_{-\pi}^{\pi}}
F_{i}^{(n)}(v_{1,2},x)\exp[\mathrm{i}mx]dx,
\end{align}
where%
\begin{align}
g(n,m,r)  &  =r^{2}-(1+m\mu)r+n^{2}\mu^{2}/4,\label{46}\\
\ \ v_{1,2}  &  =\frac{n\mu}{2r}\pm\frac{1}{r}\sqrt{g},\ \ \ a_{n}^{\prime
}(v)\ =\frac{2}{\mu}\sqrt{g};\\
n_{\max}  &  =[d(r)],\ \ \ d(r)=\frac{2(r-\sqrt{r})}{\mu}. \label{47}%
\end{align}
Here $[d]$ is the integer part of $d.$

Bringing out the distinction in the explicit form, we present $T_{i}^{(n)}$ as%

\begin{align}
T_{i}^{(n)}  &  =T_{i}^{(nr)}+T_{i}^{(ns)};\label{48}\\
T_{i}^{(nr)}  &  =(-1)^{n}\frac{\mathrm{i}}{2}\mathcal{P}%
{\textstyle\int\limits_{-1}^{1}}
dv\nonumber\\
&  \times%
{\displaystyle\int\limits_{-\pi}^{\pi}}
\left[  F_{i}^{(n)}(v,x)\frac{\exp[\mathrm{i}a_{n}(v)x]}{\sin(\pi a_{n}(v))}-%
{\displaystyle\sum\limits_{m\geq n}^{m=n_{\max}}}
{\displaystyle\sum\limits_{v_{1,2}}}
\frac{(-1)^{m}}{\pi}F_{i}^{(n)}(v_{1,2},x)\frac{\exp[\mathrm{i}mx]}%
{a_{n}(v)-m}\right]  dx,\label{49}\\
T_{i}^{(ns)}  &  =%
{\displaystyle\sum\limits_{m\geq n}^{m=n_{\max}}}
{\displaystyle\sum\limits_{v_{1,2}}}
\frac{\mu\pi}{2\sqrt{g}}\left[  1-\frac{1}{\pi}\left(  \arctan\frac{2\sqrt
{-g}}{2r-\mu n}+\arctan\frac{2\sqrt{-g}}{2r+\mu n}\right)  \right]
\label{50}\\
&  \times%
{\displaystyle\int\limits_{0}^{\pi}}
F_{i}^{(n)}(v_{1,2},x)\exp[\mathrm{i}mx]dx.
\end{align}
Here the regularized function $T_{i}^{(nr)}$ is singularity-free, and for
$n>n_{\max}\ $the integration counter in $T_{i}^{(n)}$ can be unrolled to the
lower axis. . After that we present $T_{i}$ in the form

\begin{align}
T_{i}  &  =%
{\displaystyle\sum\limits_{n>n_{\max}}^{\infty}}
T_{i}^{(n)}+%
{\displaystyle\sum\limits_{n=0}^{n_{\max}}}
T_{i}^{(n)}=\left(  T_{i}-%
{\displaystyle\sum\limits_{n=0}^{n_{\max}}}
T_{i}^{(n)}\right)  +%
{\displaystyle\sum\limits_{n=0}^{n_{\max}}}
T_{i}^{(n)}\nonumber\\
&  =%
{\textstyle\int\limits_{-1}^{1}}
dv%
{\textstyle\int\limits_{0}^{\infty}}
\left\{  F_{i}(v,x)\exp[-\chi(v,x)]+\mathrm{i}%
{\displaystyle\sum\limits_{n=0}^{n_{\max}}}
F_{i}^{(n)}(v,-\mathrm{i}x)\exp[a_{n}(v)x]\right\}  dx\nonumber\\
&  +%
{\displaystyle\sum\limits_{n=0}^{n_{\max}}}
T_{i}^{(n)}. \label{52}%
\end{align}
Here the functions $F_{i}(v,x),\chi(v,x)$ are given by Eqs.( \ref{26}),
(\ref{027}), and $a_{n}(v)$ by Eq. (\ref{40}). The integrals over $x$ in the
expression for $T_{i}^{(ns)}$ have been calculated in Appendix A \cite{[10]}.
Along with integers $m$ and $n,$ we use also $l=(m+n)/2$ and $k=(m-n)/2$ which
are straight the level numbers (see Eq. (\ref{021})). We have%

\begin{align}
\kappa_{i}^{s}  &  =\alpha m^{2}\frac{r}{\pi}%
{\displaystyle\sum\limits_{n=0}^{n_{\max}}}
\left(  1-\frac{\delta_{n0}}{2}\right)  T_{i}^{(ns)}=-\mathrm{i}\alpha
m^{2}\mu e^{-\zeta}%
{\displaystyle\sum\limits_{n,m}}
(2-\delta_{n0})\frac{\zeta^{n}k!}{\sqrt{g}l!}\nonumber\\
&  \times\left[  1-\frac{1}{\pi}\left(  \arctan\frac{2\sqrt{-g}}{2r-\mu
n}+\arctan\frac{2\sqrt{-g}}{2r+\mu n}\right)  \right]  D_{i};\ \ \ \label{53}%
\\
D_{2}  &  =\left(  \frac{m\mu}{2}-\frac{n^{2}\mu^{2}}{4r}\right)  F\nonumber\\
&  +2\mu l\vartheta(k-1)\left[  2L_{k-1}^{n+1}(\zeta)L_{k}^{n-1}(\zeta
)-L_{k}^{n}(\zeta)L_{k-1}^{n}(\zeta)\right]  ,\label{530}\\
D_{3}  &  =\left(  1+\frac{m\mu}{2}-\frac{n^{2}\mu^{2}}{4r}\right)  F+2\mu
l\vartheta(k-1)L_{k}^{n}(\zeta)L_{k-1}^{n}(\zeta),\nonumber\\
F  &  =\left[  L_{k}^{n}(\zeta)\right]  ^{2}+\vartheta(k-1)\frac{l}{k}\left[
L_{k-1}^{n}(\zeta)\right]  ^{2},\ \ \ \zeta=\frac{2r}{\mu}, \label{54}%
\end{align}
where $L_{k}^{n}(\zeta)$ is the generalized Laguerre polynomial.

At $\mu<<1,$ $(r-1)/\mu\lesssim1,$ $g/\mu\simeq|(r-1)/\mu-m|$ $<<1\ $the main
terms of sum in Eq. (\ref{53}) have a form:%

\begin{align}
\kappa_{3}^{s}  &  \simeq-\mathrm{i}\alpha m^{2}\mu e^{-\zeta}\zeta
^{m}g^{-1/2}%
{\displaystyle\sum\limits_{k+l=m}}
\frac{1}{k!l!}\label{55}\\
&  =-\mathrm{i}\alpha m^{2}\mu e^{-\zeta}\zeta^{m}g^{-1/2}\frac{2^{m}}%
{m!},\ \ \ \kappa_{2}^{s}\simeq\frac{1}{2}m\mu\kappa_{3}^{s}.
\end{align}
Here we take into account that for $\zeta>>1$%

\begin{equation}
L_{k}^{n}(\zeta)\simeq\zeta^{k}/k!,\ \ \ D_{3}\simeq\left[  L_{k}^{n}%
(\zeta)\right]  ^{2},\ \ \ D_{2}\simeq m\mu D_{3}/2. \label{56}%
\end{equation}
Eq. (\ref{55}) coincides with Eq. (\ref{13}). Note that for $g>0,$ Eq.
(\ref{55}) (as the general Eq. (\ref{53})) gives in addition the partial
probability of level population by created particles (see \cite{[10]}).

At $\mu\gtrsim1,$ $|r-1|\ <<1,$ $m=n=k=l=0,$ $D_{3}=1,$ $D_{2}=0,$ and Eq.
(\ref{53}) coincides with Eq. (\ref{25}). At $|$ $r-r_{10}|\ <<1,$ the main
term of sum is $m=n$ $=l=1,$ $k=0,$ $D_{2}=$ $\beta_{2},$ $D_{3}=$ $\beta
_{3},$ $g=-h,$ and this equation coincides with Eq. (\ref{36}).

\section{Conclusion}

So, we have investigated the photon polarization operator in weak and strong
magnetic fields for arbitrary values of the photon energy. At large quantum
numbers in a weak ($H\ll H_{0},$ $\mu\ll1$) field, there are two regions of
the photon energy. In each of these, an approximate description is of
different nature and thus a different form. The first area (with not very
large quantum numbers) is adjacent to the region of the threshold energy. From
this side it is the non-relativistic region ($r-1\ll1$). The first area
applicability ends on another side at relativistic energies ($r\gg1$), when
the parameter $\kappa=2\mu\sqrt{r}$ is not small. For these energies, SQA is
applicable, such that the energy regions of these approximations are intersect
at $\kappa\ll1$. In a weak field at $\kappa\sim1,$ the imaginary part of the
polarization operator is expressed in terms of the derivative of the Airey
function, the real part is related to the Hardy function. When $\kappa\gg1,$
the approximate description of the polarization operator is greatly simplified.

In strong fields in the expressions for the effective photon mass $\kappa_{i}%
$, we have identified integrals asymptotically diverging at the threshold
energies and taking analytically. In the remaining integrals, the contour of
integration can be moved on the imaginary axis, so that they become real
explicitly. These integrals converge well as in the integrand instead of
oscillating functions, we have exponentially falling functions. It is
necessary for the analysis of received expressions and numerical calculations.
With increasing of the photon energy, the procedure (used to the lower
threshold $r_{00}$ and $r_{10}$) could be extended in the next area to the
higher thresholds of pair creation. However, a more consistent was the
creation of the regular method to carry out the corresponding calculations in
general form. The imaginary part of the polarization operator, obtained in a
manner, coincides with the general formula for the probability of a photon
pair [10].

This work was supported in part by the Ministry of Education and Science of
the Russian Federation. The author is grateful to the Russian Foundation for
Basic Research grant №15-02-02674 for partial support of the research.


\begin{thebibliography}{99}                                                                                               %


\bibitem {[1]}M. Ruderman, in The Electromagnetic Spectrum of Nutron Stars,
NATO ANSI Proceedings (Springer, New York, 2004).

\bibitem {[2]}R.C. Duncan and C. Tompson, Astrophys. J. \textbf{392}, 19 (1992)

\bibitem {[3]}S.L. Adler, Ann. Phys. (N.Y.), \textbf{67}, 599 (1971).

\bibitem {[4]}J. Schwinger, Phys. Rev., \textbf{82, }664\textbf{
}(1951)\textbf{.}

\bibitem {[5]}I.A. Batalin and A.E.Shabad, Sov. Phis. JETP \textbf{33}, 483 (1971).

\bibitem {[6]}V.N. Baier, V.M. Katkov and V.M. Strakhovenko, Sov. Phis. JETP
\textbf{41}, 198 (1975).

\bibitem {[7]}V.N. Baier, A.I. Milstein and R.Zh. Shaisultanov, Zh. Eksp.
Teor. Fiz. \textbf{111}, 52 (1997).

\bibitem {[8]}A.C. Harding, M.G. Baring and P.L. Conthier, $\ $Astrophys. J.
\textbf{476}, 246 (1997).

\bibitem {[9]}A.E.Shabad, Zh. Eksp. Teor. Fiz. \textbf{125}, 210, (2004).

\bibitem {[10]}V.N. Baier and V.M. Katkov, Phys. Rev., D \textbf{75}, 073009 (2007).

\bibitem {[11]}V.M. Katkov, Sov. Phis. JETP \textbf{141}, 258 (2012).
\end{thebibliography}
\end{document}